\documentclass[prb,twocolumn,unsortedaddress,altaffilletter,preprintnumbers,amsmath,amssymb,superscriptaddress]{revtex4}

\usepackage{graphicx}
\usepackage{bm}
\usepackage{color}
\usepackage{amsmath}
\usepackage{amssymb}

\begin{document}

\title{Detection of bottom ferromagnetic electrode oxidation in magnetic tunnel junctions by magnetometry measurements}

\author {Wei Chen}
\thanks{Corresponding author}
\email{wc6e@virginia.edu}
\affiliation{Department of Physics, University of Virginia, Charlottesville, VA 22904, USA}

\author {Dao N. H. Nam}
\email{daonhnam@yahoo.com}
\affiliation{Department of Materials Science and Engineering, University of Virginia, Charlottesville, VA 22904, USA}

\author {Jiwei Lu}
\affiliation{Department of Materials Science and Engineering, University of Virginia, Charlottesville, VA 22904, USA}

\author {Stuart A. Wolf}
\affiliation{Department of Physics, University of Virginia, Charlottesville, VA 22904, USA}
\affiliation{Department of Materials Science and Engineering, University of Virginia, Charlottesville, VA 22904, USA}
\date{\today}
\begin{abstract}
Surface oxidation of the bottom ferromagnetic (FM) electrode, one of the major detrimental factors to the performance of a Magnetic Tunnel Junction (MTJ), is difficult to avoid during the fabrication process of the MTJ's tunnel barrier. Since Co rich alloys are commonly used for the FM electrodes in MTJs, over-oxidation of the tunnel barrier results in the formation of a CoO antiferromagnetic (AF) interface layer which couples with the bottom FM electrode to form a typical AF/FM exchange bias (EB) system. In this work, surface oxidation of the CoFe and CoFeB bottom electrodes was detected via magnetometry measurements of exchange-bias characterizations including the EB field, training effect, uncompensated spin density, and coercivity. Variations of these parameters were found to be related to the surface oxidation of the bottom electrode, among them the change of coercivity is most sensitive. Annealed samples show evidence for an oxygen migration back to the MgO tunnel barrier by annealing.
\end{abstract}
%
%
\maketitle
\section{Introduction}
Since the discovery of room temperature (RT) large tunnel magnetoresistance (TMR) in magnetic tunnel junctions (MTJs),\cite{1,2} intensive
research has been carried out on this subject. This is due in large part to the various potential technological applications utilizing the high
TMR performance,\cite{3} which is quickly approaching the theoretically predicted value of over 1000\% in Fe/MgO/Fe MTJ structure.\cite{4}
Recently, a TMR value of 605\% was reported in a pseudo spin valve annealed at high temperature \cite{5} and a record high TMR value of 1056\%
in a double barrier MTJ structure.\cite{6} However, the performance of MTJs is well known to be very sensitive to fabrication conditions, and
significant challenges remain for fabrication of reliable high quality MTJs. Among those challenges, one critical step is the formation of oxide
tunnel barrier in between the two FM metallic electrodes. The oxidation step needs precise control due to the small thickness of the barrier,
which is usually less than 2 nm, and the unwanted surface oxidation of the bottom FM electrode. If the barrier is under-oxidized, the
leftover metal could significantly diminish the tunneling by providing hopping conductance for electrons within the barrier. On the other
hand, if the barrier is over-oxidized, oxidation of the ferromagnetic (FM) electrode could occur and it is severely detrimental to the spin polarization at the electrode/barrier interface.\cite{7} In the current stage of development, most ferromagnetic electrodes used in the MTJ structures are Co rich alloys such as CoFe and CoFeB. In the case of barrier over-oxidation, the bottom electrode surface will be oxidized to form antiferromagnetic CoO at the bottom electrode/barrier interface. When such a system is field-cooled from RT through the N\'{e}el temperature of
CoO (293 K), a typical CoO/FM exchange-bias (EB) system is established.\cite{8} In this study, the EB system is utilized as an indicator of
barrier over-oxidation. Specifically, we identify four parameters associated with the observation of EB as the detection parameters of barrier
over-oxidation, including the EB field ($H_{eb}$), training effect, uncompensated spin density, and coercivity ($H_c$) of the bottom electrode.
A comparison of the detection sensitivities for these four parameters shows that $H_c$ is most sensitive to the over-oxidation due to its very
pronounced increase at low temperature. Our study also shows that annealing at high temperature would help an oxygen remigration from the
electrode oxidized surface to the MgO barrier, in agreement with previously reported results obtained by advanced spectroscopic methods.\cite{9}
We propose that the magnetic measurements and analyses developed in this study can serve as a simple and effective method for detecting the
tunnel barrier oxidation quality during the MTJ fabrication process. In particular, it will be a useful tool for identifying the process windows
when exploring new oxide barrier materials other than the well established Al$_2$O$_3$ and MgO.
\section{Experiment}
\begin{figure}[]
  \includegraphics[width=2.5in, keepaspectratio=true, angle=0]{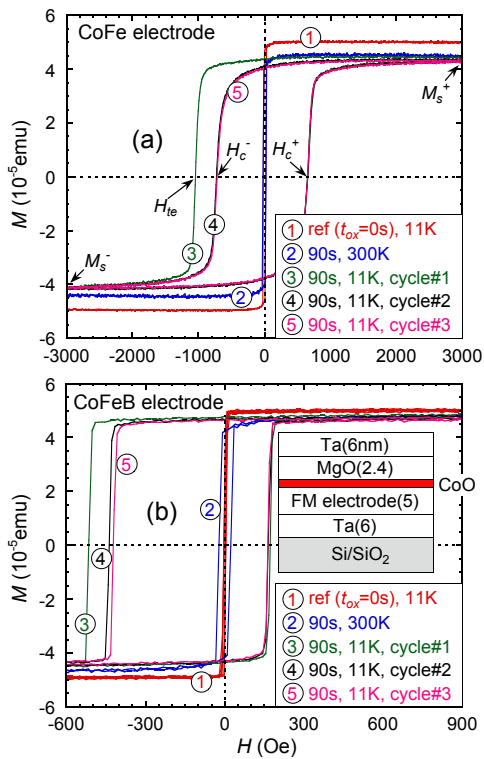}
  \caption{(Color online) RT and 11 K $M$($H$) loops of the samples with (a) CoFe and (b) CoFeB bottom electrode. For all the oxidation samples, the Mg layers were oxidized in the same condition for 90 seconds. The loops (1) for the corresponding reference (un-oxidized) samples at 11 K were also presented. The inset of (b) shows the general structure of the partial MTJ stacks used in this study.}\label{Fig1}
\end{figure}
The partial MTJ stacks used in this study were deposited using a biased target ion beam deposition (BTIBD) system \cite{10,11,12} with the base
pressure of $\sim$2$\times$10$^{-7}$ Torr and the processing pressure of $\sim$7$\times$10$^{-4}$ Torr (equivalent to 80 sccm Ar flow during deposition). An initial magnetic easy axis of the FM layers was set by a magnetic field of 50 Oe applied parallel to the plane of the film in situ during the film growth. The studied partial MTJ structure, as shown in the inset of Fig. 1 (b), is: substrate/Ta(6)/FM(5)/MgO(2.4, post-oxidation)/Ta(6); here, the thickness units are in nm, the substrate is thermally oxidized silicon, and the bottom FM electrode material is either Co$_{95}$Fe$_5$ or Co$_{60}$Fe$_{20}$B$_{20}$. The MgO layer was formed by an 80 sccm Ar/3 sccm O$_2$ plasma oxidation following the deposition of a 2.4 nm Mg metallic film. For each of the two FM electrode materials, samples with varying oxidation times, $t_{ox}$, were fabricated to study the sensitivity of different detection parameters. The reference samples, substrate/Ta(6)/FM(5)/Mg(2.4)/Ta(6), were fabricated with similar structures as the oxidized ones except without a plasma oxidation step (i.e., $t_{ox} = 0$ s). To investigate the influence of annealing on the magnetic properties, the samples were annealed at two different temperatures (200 $^\mathrm{o}$C and 350 $^\mathrm{o}$C) in an external field of 3 kOe applied parallel to the film plane. The samples were entirely protected from further oxidation during annealing by a continuously flowing forming gas (95\% N$_2$ + 5\% H$_2$). Magnetic properties were measured using a Quantum Design PPMS-6000 system at both room temperature (RT) and 11 K. For the low temperature measurements, all of the samples were field-cooled from RT with an applied magnetic field of 3 kOe along their easy axes. The hysteresis loops of each sample were consecutively repeated three times at 11 K in order to observe the training effect associated with the exchange bias phenomenon.

\section{Results and Discussion}
\begin{figure}[t!]
  \includegraphics[width=2.2in]{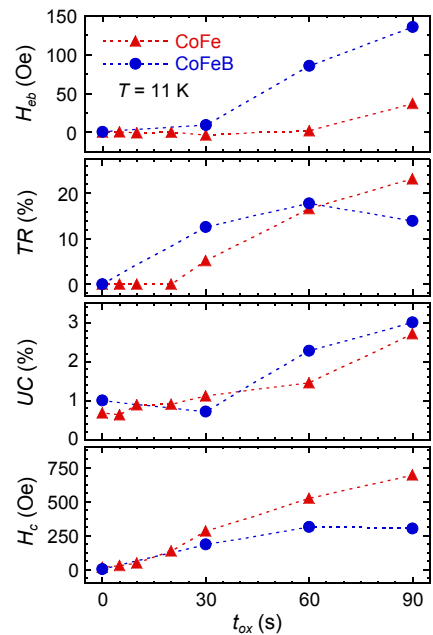}
  \caption{(Color online) The oxidation time dependence of the four detection parameters $H_{eb}$, $TR$, $UC$ and $H_c$ measured at 11 K for both CoFe- ($\blacktriangle$) and CoFeB-electrode ($\bullet$) samples after a field cooling in $H = 3$ kOe from RT.}\label{Fig2}
\end{figure}
\subsection{Parameters associated with EB effect}
As mentioned in the introduction part, when the bottom FM electrode (CoFe or CoFeB) gets surface oxidized because of an over-oxidation of the MgO barrier, a typical CoO/FM EB system is formed at low temperature following a field cooling through 293 K. Fig. \ref{Fig1} shows the magnetic hysteresis loops of two representative samples with bottom FM electrodes of CoFe [Fig. \ref{Fig1}(a)] and CoFeB [(Fig. \ref{Fig1}(b)], respectively. The plasma oxidation time $t_{ox}$ is 90 s for both samples. Also shown in Fig. \ref{Fig1} are the loops from the corresponding reference samples that have the same structure but with no oxidation step carried out for the Mg metal layers. The reference samples exhibit no change in the $M$($H$) loops by a multiple field cycling and a very small coercivity ($<$20 Oe) even at 11 K. Similar results, which are omitted for clarity, were also observed at RT. Although the oxidized samples show no unusual $M$($H$) loops at RT, the low temperature measurements exhibit a clear training effect and dramatic increase in coercivity compared to the reference ones. It is noticeable that the first cycling irregular shaped loop with extremely large $H_{eb}$ and $H_c$ is known as due to the training effect of typical EB systems.\cite{13} The system becomes stable only after the second field cycle illustrated by a near coincidence of the second and third cycling loops. To quantify the training effect, we define a "training ratio" ($TR$) that describes the magnitude of the effect: $TR(\%)=[(H_{te}-H_c)/H_c]\times100\%$, where $H_{te}$ is the training field that is determined as the coercivity measured on the first demagnetization branch of the first cycling $M$($H$) loop as indicated in Fig. \ref{Fig1}. After the second field cycle, the training effect almost ends and the $M$($H$) loops go back to the regular shape where the EB field is conventionally determined as $H_{eb}=\frac{1}{2}(H_c^++H_c^-)$, and the coercivity as $H_{c}=\frac{1}{2}(H_c^+-H_c^-)$, with $H_c^+$ and $H_c^-$ are the positive and negative fields, respectively, at which $M = 0$. The variations of $H_{eb}$, $H_c$ and $TR$ with the plasma oxidation time are presented in Fig. \ref{Fig2}.

It has been generally believed that highly anisotropic uncompensated spins at the FM/AFM interface are responsible for unidirectionally pinning
the FM layer.\cite{14} The net magnetization of uncompensated spins, in principle, causes a vertical shift along the $M$-axis to the hysteresis
loop. The density of the uncompensated spin moments can be determined as $UC(\%)=[(M_s^+-M_s^-)/(M_s^++M_s^-)]\times100\%$, where $M_s^+$ and
$M_s^-$ are the saturation magnetization of the samples at positive and negative fields, respectively. In conventional exchange bias systems,
$UC$ is usually very small and the shift caused to the $M$($H$) loop is hardly observed. Interestingly, $UC$ is quite obvious in our samples with
long oxidation times. The variations of $UC$ vs. $t_{ox}$ for all of the samples are also plotted in Fig. \ref{Fig2}.
\begin{figure}[t!]
  \includegraphics[width=2.2in]{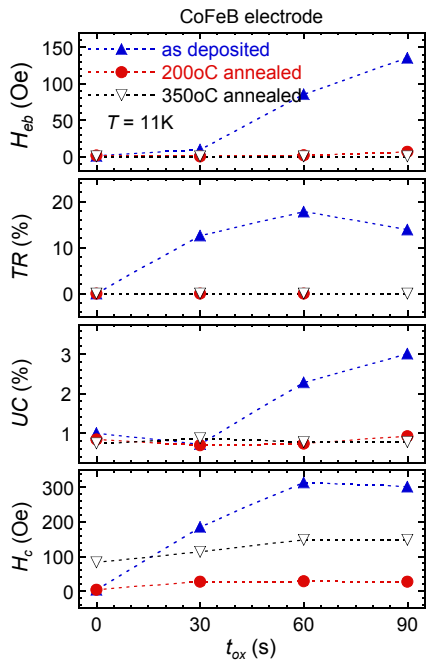}
  \caption{(Color online) Effect of annealing on the detection parameters for CoFeB bottom electrode samples. Up triangles, circles, and down triangles are for the as-deposited, 200 $^\mathrm{o}$C, and 350 $^\mathrm{o}$C annealed samples, respectively. All measurements were performed at 11 K.}\label{Fig3}
\end{figure}

The results in Fig. \ref{Fig2} show a clear trend for all the parameters: almost no change for very short oxidation times, but substantial increases in all of the parameters with sufficiently prolonged $t_{ox}$. These variations of the parameters are therefore clear evidences for the formation of the CoFe/CoO or CoFeB/CoO interface caused by the barrier over-oxidation at long $t_{ox}$. For each considered parameter, the oxidation time threshold where it starts to increase can be considered as the boundary between under- and over-oxidation of the Mg barrier. Our observations clearly show that all the four EB related parameters presented in our study can be employed effectively for detecting the bottom FM electrode surface oxidation. The oxidation time threshold slightly varies among these parameters, indicating that they may have different sensitivities with the bottom electrode oxidation.

To detect the barrier layer over-oxidation that causes the bottom electrode to be partially oxidized, one apparent approach would be to compare the saturation magnetization of the oxidized sample to the reference one. If the surface of the FM bottom electrode gets oxidized to form non-FM oxides, its total FM moment must be reduced. Indeed, a reduced magnetic moment of the sample with over-oxidation of MgO layer [Fig. \ref{Fig1}(a)] was observed when compared to the reference sample loop measured at RT (not shown). However, this parameter of reduced moment is not pursued in this study due to the significant inaccuracy in normalizing the film's surface area and the fact that this parameter is not directly related to the EB phenomenon.

\subsection{Sensitivity of the EB parameters}

The EB parameters appear to have different sensitivities with the oxidation (see Fig. \ref{Fig2}). For the sample set with CoFe electrodes, when decreasing oxidation time from 90 seconds, the $H_{eb}$ of corresponding samples starts to drop quickly and becomes almost zero on the sample with $t_{ox} = 30$ s. The training effect almost disappears in the sample with $t_{ox} = 20$ s while $UC$ is still measurable even with only 10 s oxidation. Coercivity seems to be the most sensitive parameter; an enhanced coercivity (comparing to the $t_{ox} = 0$ s reference sample) due to the EB could still be clearly observed even with only 5 seconds of oxidation. This enhanced coercivity indicates that the MgO barrier was still slightly over-oxidized (even with only 5 s oxidation), leading to the surface oxidation of the bottom CoFe electrode at the MgO/CoFe interface and the formation of the CoO/CoFe EB system. Qualitatively, the data from the CoFeB-electrode oxidized samples present a similar trend in terms of detection sensitivities among those four parameters. The results from both sets of samples show that the coercivity is the most sensitive parameter to detect the presence of EB formation, followed by the uncompensated spin density $UC$, the training ratio $TR$, and finally the EB field $H_{eb}$. By measuring these parameters, one can quickly examine the oxidation state of the CoFeB or any other Co rich FM electrode underneath the oxide barrier, providing the optimization of the tunnel barrier oxidation with simplicity without the characterization of the completed MTJ structure with microfabrication process or other sophisticated techniques.

\subsection{Annealing effect on detection parameters}
\begin{figure}[t!]
  \includegraphics[width=2.2in]{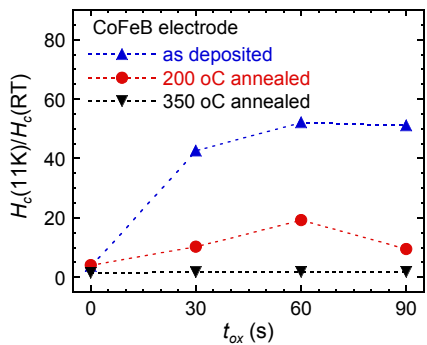}
  \caption{(Color online) The normalized coercivity $H_c$(11K)/$H_c$(RT) for as-deposited and annealed samples as functions of plasma oxidation time exhibiting a gradual decrease with increasing the annealing temperature.}\label{Fig4}
\end{figure}

For MTJs with MgO barrier, it is necessary to anneal the samples in order to obtain crystallized MgO(001) for high spin filtering efficiency.\cite{15} Previous study of CoFeB/MgO bilayer using x-ray photoemission spectroscopy (XPS) not only confirmed the process-dependent formation of CoO at the CoFeB/MgO interface, which is similar to our findings using this over-oxidation detection method, but also unveiled the reduction of such oxidation during the vacuum annealing process.\cite{9} To verify the existence of similar oxidation reduction in our samples using the EB detection parameters after the heat treatment, the sample set with CoFeB bottom electrode, including both the reference sample and the oxidized samples, was annealed for one hour at two different temperatures 200 $^\mathrm{o}$C and 350 $^\mathrm{o}$C. Magnetometry measurements were then performed on the annealed samples as well as on those as-deposited ones and the obtained results are summarized in Fig. \ref{Fig3}. After the annealing treatments, both at 200 $^\mathrm{o}$C and 350 $^\mathrm{o}$C, except the $H_c$, all other parameters become almost the same as those of the reference sample. This observed reduction of $H_{eb}$, $TR$ and $UC$ values serves as a clear indication of the deoxidation state of the bottom CoFeB electrode after the annealing, which suggests that during the annealing process, oxygen atoms at the CoFeB oxidized surface may have diffused back to the MgO barrier, decreasing the amount of CoO and therefore weakening the CoO/CoFeB exchange coupling. This result is in good agreement with previous studies using XPS and similar magnetometry measurements.\cite{9,16}

One thing worth noting here is the larger $H_c$ of the samples annealed at 350 $^\mathrm{o}$C compared to those at 200 $^\mathrm{o}$C. This coercivity enhancement is probably due to the crystallization of the CoFeB film; it obscures the annealing effect on the underlying EB coupling and the oxidation state at the interface. To further clarify the influence of annealing on the coercivity, it's normalized value, $H_c$(11K)/$H_c$(RT), is plotted against $t_{ox}$ in Fig. \ref{Fig4}. The result clearly shows that the CoFeB surface oxidation is not completely reversed after the one hour annealing at 200 $^\mathrm{o}$C, and the EB coupling still exists in all of the oxidation samples. However, after the annealing at 350 $^\mathrm{o}$C, all the oxidation samples show almost identical $H_c$ as the reference one, indicating a disappearance of the CoFeB surface oxidation and therefore the exchange bias when most of the oxygen atoms have been absorbed by the MgO barrier. This conclusion is consistent with the annealing study of MTJs with MgO barriers where it was found that, within a certain temperature range, higher temperature annealing usually yields better TMR performances.\cite{17} In addition, the annealing effect study also confirms that among the four detection parameters, coercivity $H_c$ is the most sensitive probe for detecting the bottom electrode oxidation in the MTJ fabrication process.

\section{Conclusion}
In summary, we have developed a simple method of detecting the surface oxidation of the bottom FM electrode due to an over-oxidation of the MgO
barrier layer that can be used as a generic detection method to explore the oxidation process window of various barrier materials regardless of
their types. Four different oxidation detection parameters were identified, including the newly defined training ratio, $TR$, which
can all be used to monitor the exchange bias formation. While the exchange bias field $H_{eb}$ itself is not detectable at short oxidation times,
the enhanced coercivity due to the exchange bias formation appears to be most sensitive. Our study of the annealing effect on these four
parameters confirmed the oxygen restoration to the barrier material, which is consistent with the results from previous studies using different
experimental techniques.

\begin{acknowledgments}
This work has been performed with financial support from UCR/DMEA under Grant No. H94003-08-2-0803, ONR/MURI under Grant No. N00014-06-1-0428,
and ARO/MURI under Grant No. W911NF-08-2-0032. Many helpful discussions and comments from Dr. W. F. Egelhoff, Jr. (NIST) are highly acknowledged.
\end{acknowledgments}

\end{document}